          [1999/12/02 1.01 (PWD)]
\documentclass[a4paper,twocolumn]{esapub} 
\usepackage{natbib}
\usepackage{graphicx}

\title{A SAMPLE OF RADIUS-EXPANSION BURSTERS OBSERVED BY THE BeppoSAX-WFCs}
\author{Massimo Cocchi}
\author{A. Bazzano}
\author{L. Natalucci}
\author{P. Ubertini}
\affil{Istituto di Astrofisica Spaziale (IAS/CNR), via Fosso del Cavaliere, 00133 Roma, Italy}
\author{J. Heise}
\author{E. Kuulkers}
\author{M.J.S. Smith}
\author{J.J.M. in 't Zand}
\affil{Space Research Organization Netherlands (SRON), Sorbonnelaan 2, 3584 CA, Utrecht, The Netherlands}

\begin{document}


\maketitle

\begin{abstract}
Since 1996 a monitoring campaign of the Galactic Bulge region has been performed by the {\em BeppoSAX} Wide 
Field Cameras in the 2--28 keV band.  This allowed accurate and unbiased studies of classes of objects. In 
particular the type-I X-ray bursters, whose distribution is concentrated in the monitored sky region, have 
been investigated. High luminosity radius-expansion bursts have been observed from nine sources, allowing 
to estimate important physical parameters such as distance, photospheric radius, averaged luminosity and 
accretion rate. An overview of these results is presented.
\end{abstract}

\section{Introduction}

Following the classification proposed by Hoffman (1978), X-ray bursts are divided in two main types: type-I, type-II 
(see Lewin, van Paradijs \& Taam, 1993 for a comprehensive review).
While type-II bursts have been observed only in two peculiar cases, namely the {\em Rapid Burster} (MXB~1730$-$335) and the 
{\em Bursting Pulsar} (GRO~J1744$-$28), type-I bursts are much more common and likely originate from thermonuclear flashes 
onto a neutron star surface. 
All the type-I bursting sources are associated with low-mass binary systems (LMXBs), following the identification of their 
optical counterparts or, indirectly, from spectral and/or stellar population (e.g. globular cluster sources) 
characteristics. \\  
It is not uncommon to observe type-I bursts so energetic as to reach the Eddington luminosity for a typical 
neutron star ($\sim 2\times 10^{38} {\rm erg~s}^{-1}$). 
During such events the atmosphere of the neutron star adiabatically expands due to radiation pressure.  
In the expansion phase the burst luminosity is almost constant and very close to the Eddington limit. 
The blackbody temperature decreases during the expansion, and in some cases it gets so low that X-rays are no longer emitted, 
thus originating a {\em precursor} event (Tawara et al., 1984; Lewin, Vacca \& Basinska, 1984).  
The subsequent decrease of the photosphere radius down to its original value causes the temperature to increase and harder 
X-rays to be emitted again, until the burst emission eventually cools down. \\
In most cases the radius increase is not so large as to generate a precursor, but the temperature shifts during the 
expansion-contraction phases cause a double peak to be observed at high photon energies (typically above 5 keV).  
Since the peak luminosity of such energetic events is Eddington-limited, radius expansion bursts could be regarded as 
standard candles, assuming the bursts to radiate isotropically and the Eddington luminosity to be the same for all the 
bursting sources (e.g. Lewin, van Paradijs \& Taam, 1993). \\
The distances inferred from the observations of Eddington-limited bursts are of help in the study of neutron-star binaries, 
allowing to estimate neutron star radii, average luminosities and accretion rates.  
During $\sim 4$ years of BeppoSAX in-flight operation, at least 10 sources showing radius-expansion bursting behaviour were 
observed by the WFCs.  An overview of the results obtained on a sample of 9 of them is presented in the next section.

\begin{figure}[htb]
\centering
\includegraphics[width=1.\linewidth]{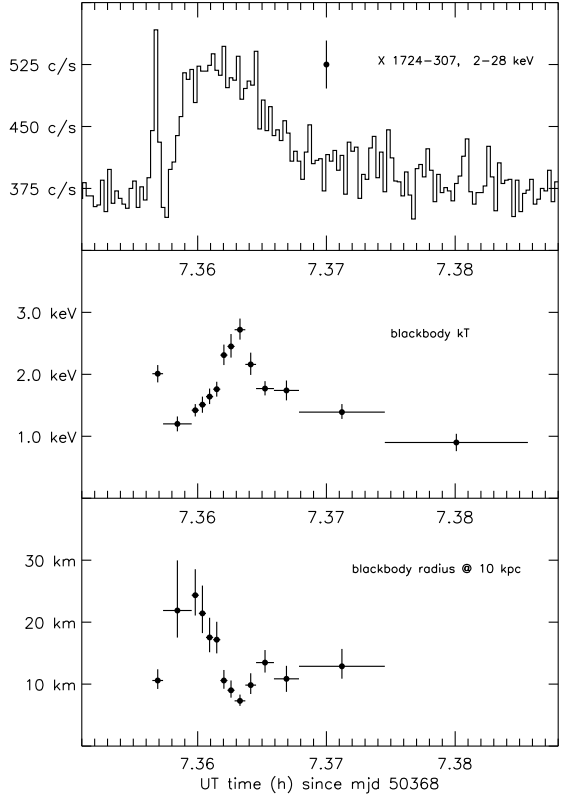}
\caption{An example of Eddington-limited burst with precursor event: 1E 1724-3045.\label{fig:double}}
\end{figure}

\begin{figure}[htb]
\centering
\includegraphics[width=1.\linewidth]{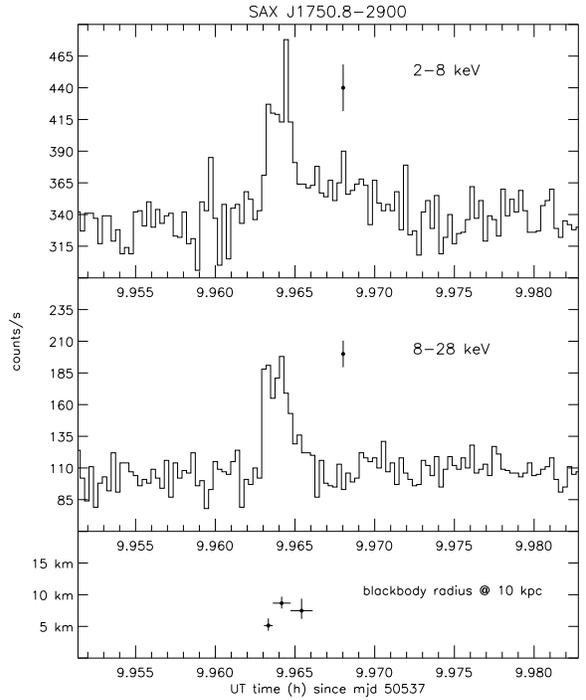}
\caption{The radius-expansion burst observed from SAX J1750.8-2900 in Spring 1997.\label{fig:single}}
\end{figure}

\begin{figure}[htb]
\centering
\includegraphics[width=1.\linewidth]{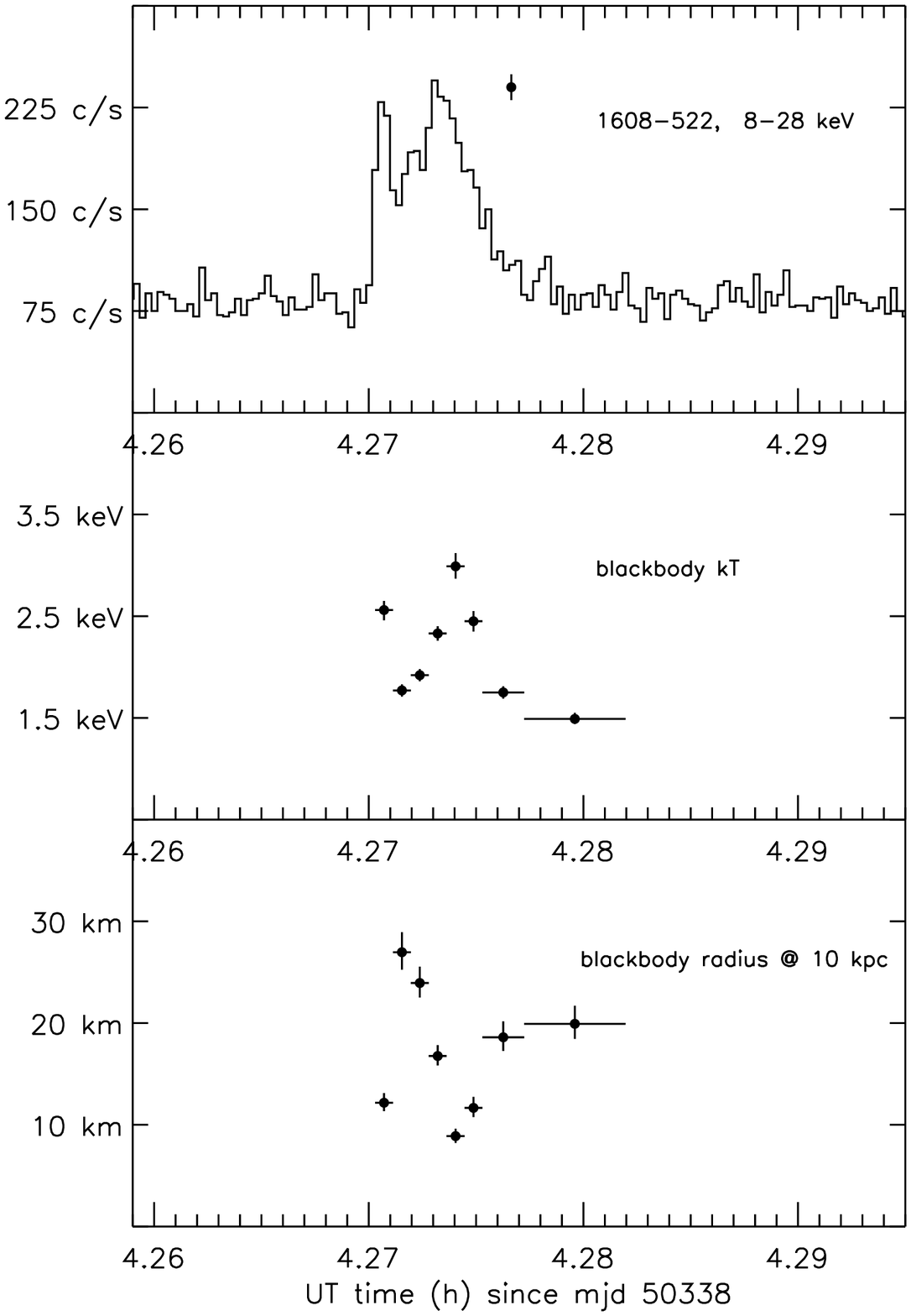}
\caption{A radius-expansion burst from 1608-522.\label{fig:single}}
\end{figure}

\section{Observation and data analysis}

The Wide Field Cameras (WFC) on board the {\em BeppoSAX} satellite consist of two identical coded mask telescopes 
(Jager et al., 1997).  
The two cameras point at opposite directions each covering $40^{\circ}\times 40^{\circ}$ field of view. 
With their source location accuracy in the range $1-3'$, a time resolution of 0.244 ms at best, and an energy resolution 
of 18\% at 6 keV, the WFCs are very effective in studying hard X-ray (2--28 keV) transient phenomena.  
The imaging capability, combined with the good instrument sensitivity (5--10 mCrab in $10^{4}$ s), allows accurate 
long term monitoring of complex sky regions like the Galactic Bulge (see Heise et al., 1999, and Ubertini et al., 1999, 
for a review of the WFC monitoring campaign of the Galactic centre region). \\
The data of the two cameras is systematically searched for bursts or flares by analysing the time profiles of the detector 
in the 2--11 keV energy range with time resolution down to 1~s.  
Reconstructed sky images are generated in coincidence with any statistically meaningful enhancement, to identify possible 
bursting events.  The accuracy of the reconstructed position, which of course depends on the intensity of the burst, is 
typically better than $5'$.
This analysis procedure allowed for the identification of more than 1400 X-ray bursts from 35 type-I bursters in a total of 
$\sim 4$ Ms WFC net observing time (see in 't Zand, these proc., for an updated review). \\
A total of 10 sources, 5 of which newly discovered by the BeppoSAX-WFCs (among them, the accreting millisecond pulsar 
SAX~J1808.4$-$3658) and two detected bursting for the first time (GRS~1741.9$-$2853 and 1RXS~J171824.2-402934), were observed 
to produce Eddington-limited bursts.
This increases the number of the known radius-expansion bursters by $\sim 40\%$. 
In this paper we analyse a sample of 9 sources, not including the long-duration ($>200$ s) burst observed from 
1RXS~J171824.2-402934 (see Kaptein et al., 2000, for details on this rather atypical burster). 
In the analysed sample, only 1E~1724$-$3045 in Terzan 2 showed clear evidence for precursor events (Figure 1), indicating 
strong expansion of its photosphere.  
The other sources exhibited evident double-peaked profiles at high energies ($> 8$ keV).
Examples of Eddington-limited burst profiles observed by the WFCs are given in Figure 1--3.
In particular, in Figure 2 the time profile of the only radius-expansion burst detected from the 1997 transient 
SAX~J1750.8$-$2900 (Natalucci et al. 1999) is shown. \\
In Table 1 the results obtained for the radius-expansion bursters in our sample are summarised. 
The distances are calculated assuming isotropic emission and an Eddington luminosity of $2.5\times 10^{38} {\rm erg s}^{-1}$, 
which is appropriate for a $1.4 M_{\odot}$ neutron strar and helium-rich fuel.  This value also includes a moderate (20\%)
gravitational redshift correction (van Paradijs \& Mc Clintok, 1994).
We did not correct to effective blackbody temperature from color temperature (see Lewin, van Paradijs \& Taam, 1993, for details).  
The observed stellar radii $R_{\infty}$ are all in the range $\sim 4-7$ km, and their average value is 5.5 km. 
The tabulated errors for the distance and $R_{\infty}$ are $1 \sigma$, but the systematic uncertainties associated with the 
assumption of standard burst parameters are of course larger. \\
Our sample consists of six transients (the five discovered by the WFCs plus GRS~1741.9$-$2853) and three relatively bright 
persistent sources (1608$-$522, 1E~1724$-$3045 and 4U~1812$-$12).     We found the bolometric luminosity of the 
transients not to exceed $\sim 10^{36} {\rm erg~s}^{-1}$ during quiescence, and their outburst peak luminosity to be well below the 
Eddington limit (within $\sim 30\%$).  
The average luminosities of the three persistent sources are in excess of $4\times 10^{36} {\rm erg~s}^{-1}$.
In general, we observed bright X-ray bursts when the source intensity was lower than average, i.e. in low state for the persistent 
emitters, or in the decay phase after the outburst, for the transients.  This is in agreement with previous studies (see 
Lewin, van Paradijs \& Taam, 1993).
For each source in the analysed sample, the observed parameters are briefly discussed hereafter.

\begin{table*}
  \begin{center}
    \caption{Eddington-limited type-I X-ray bursters observed by the BeppoSAX-WFC in 1996-2000.}\vspace{1em}
    \renewcommand{\arraystretch}{1.2}
    \begin{tabular}[htb]{lcccl}
      \hline
      source  & 
      Peak Flux  &  
      Distance &
      $R_{\infty}$ &  
      WFC reference \\
      & ($10^{-8} {\rm erg~cm}^{-2} {\rm s}^{-1}$) & (kpc) & (km) & \\
      \hline
      1608$-$522          &  $15.7\pm 1.6$  &  $3.6\pm 0.2$  &  $4.2\pm 0.4$  &  Smith et al. (1999) \\
      SAX~J1712.6$-$3739  &  $ 6.0\pm 1.0$  &  $5.9\pm 0.5$  &  $5.3\pm 1.2$  &  Cocchi et al. (1999a; 2000b) \\
      1E~1724$-$3045      &  $ 4.1\pm 0.4$  &  $7.1\pm 0.3$  &  $6.7\pm 0.7$  &  Cocchi et al. (1999b) \\
      GRS~1741.9$-$2853   &  $ 2.9\pm 0.4$  &  $8.5\pm 0.6$  &  $6.0\pm 1.1$  &  Cocchi et al. (1999c) \\
      SAX~J1747.0$-$2853  &  $ 3.8\pm 0.6$  &  $7.4\pm 0.6$  &  $4.9\pm 0.9$  &  Natalucci et al. (2000b) \\
      SAX~J1750.8$-$2900  &  $ 3.7\pm 0.5$  &  $7.5\pm 0.5$  &  $5.4\pm 0.8$  &  Natalucci et al. (1999) \\
      SAX~J1808.4$-$3658  &  $26.1\pm 2.0$  &  $2.8\pm 0.1$  &  $6.1\pm 0.8$  &  In 't Zand et al. (1998; 2000) \\
      SAX~J1810.8$-$2609  &  $ 9.2\pm 1.5$  &  $4.8\pm 0.4$  &  $6.7\pm 1.1$  &  Natalucci et al. (2000a) \\
      4U~1812$-$12        &  $18.0\pm 1.6$  &  $3.4\pm 0.1$  &  $5.8\pm 0.3$  &  Cocchi et al. (2000a) \\
      \hline \\
      \end{tabular}
    \label{tab:table}
  \end{center}
\end{table*}

\subsection{1608$-$522}
This source is a well-known X-ray burster. 
Average bolometric intensities of $\sim 2\times 10^{-8} {\rm erg~cm}^{-2} {\rm s}^{-1}$ and 
$\sim 5\times 10^{-9} {\rm erg~cm}^{-2} {\rm s}^{-1}$ were observed by {\em Tenma} for the high and low source state, 
respectively (Nakamura et al., 1989).
For the determined 3.6 kpc distance, luminosities of $\sim 3\times 10^{37} {\rm erg~s}^{-1}$ and 
$\sim 8\times 10^{36} {\rm erg~s}^{-1}$ are obtained for the two states, corresponding to 15\% and 4\% of the Eddington 
luminosity.  At the time of the burst observed by the WFCs (Fig. 3), the source was not detected with a 2--28 keV $3\sigma$ upper 
limit of $2\times 10^{-9} {\rm erg~cm}^{-2} {\rm s}^{-1}$, consistent with the low-state average intensity.
Assuming standard neutron star parameters ($M = 1.4 M_{\odot}, R = 10$ km), the 
average accretion rate can be extrapolated in the two luminosity states.  We obtain average rates of 
$\sim 3\times 10^{-9} M_{\odot} {\rm y}^{-1}$ ($6\times 10^{4} {\rm g~cm}^{-2} {\rm s}^{-1}$) and 
$\sim 7\times 10^{-10} M_{\odot} {\rm y}^{-1}$ ($1\times 10^{4} {\rm g~cm}^{-2} {\rm s}^{-1}$) for the high and 
the low state, respectively. 

\subsection{SAX~J1712.6$-$3739}
The source, discovered by the WFCs in late Summer 1999 (in 't Zand et al., 1999), showed a single type-I bursting event so far 
(Cocchi et al., 1999a).  
During the outburst, its X-ray intensity peaked at 32 mCrab in the 2--10 keV bandpass, corresponding to 
$\sim 6.5\times 10^{-10} {\rm erg~cm}^{-2}{\rm s}^{-1}$.  At the time of the burst detection the source was below the 
typical WFC $3 \sigma$ sensitivity level, i.e. $2\times 10^{-9} {\rm erg~cm}^{-2}{\rm s}^{-1}$ (2--28 keV).
This exptrapolates, taking into account the spectrum observed by the {\em BeppoSAX}-NFIs (Cocchi et al. 2000b) and the $\sim 6$ 
kpc distance, to a bolometric luminosity of $\sim 10^{36} {\rm erg~s}^{-1}$.  During the outburst, the luminosity was 
$\sim 2\times 10^{37} {\rm erg~s}^{-1}$, $\sim 10\%$ of the Eddington limit for a standard neutron star.
The inferred mass accretion rates are $< 10^{-10} M_{\odot} {\rm y}^{-1}$ ($2\times 10^{3} {\rm g~cm}^{-2} {\rm s}^{-1}$) and 
$1.6\times 10^{-9} M_{\odot} {\rm y}^{-1}$ ($3.4\times 10^{4} {\rm g~cm}^{-2} {\rm s}^{-1}$) for the quiescent and the outburst phases, 
respectively.

\subsection{1E~1724$-$3045}
This well-studied bright persistent source, located in the globular cluster Terzan 2, is a typical example of Eddington-limited  
burster with precursor events (see Fig. 1).  All the 14 bursts observed by the {\em BeppoSAX}-WFCs showed this feature, indicating 
the bursts of 1E~1724$-$3045 are very luminous and cause strong expansion of the photosphere. For the estimated 7.1 kpc distance, 
bolometric luminosities in the range $1-3\times 10^{37} {\rm erg~s}^{-1}$ are generally observed (e.g. Olive et al., 2000).  
This, for standard neutron star parameters, leads to accretion rate values in the range $1-3\times 10^{-9} M_{\odot} {\rm y}^{-1}$, 
corresponding to $2-6\times 10^{4} {\rm g~cm}^{-2} {\rm s}^{-1}$.

\subsection{GRS~1741.9$-$2853}
During the only outburst reported so far, observed by {\em GRANAT} in 1990, the 4--20 keV intensity reached 9.6 mCrab 
(e.g. Pavlinsky et al., 1994).  Taking into account the observed spectrum and the distance inferred by the  
Eddington-limited burst observed by the {\em BeppoSAX}-WFCs in 1996 (Cocchi et al., 1999c), we extrapolate an outburst bolometric 
luminosity of $\sim 5\times 10^{36} {\rm erg~s}^{-1}$.  At the time of the burst detection, a $3 \sigma$ upper limit of 
$\sim 2\times 10^{36} {\rm erg~s}^{-1}$ was obtained by the WFCs.  The corresponding accretion rates are 
$4\times 10^{-10} M_{\odot} {\rm y}^{-1}$ ($8\times 10^{3} {\rm g~cm}^{-2} {\rm s}^{-1}$) and 
$< 2\times 10^{-10} M_{\odot} {\rm y}^{-1}$ ($< 4\times 10^{3} {\rm g~cm}^{-2} {\rm s}^{-1}$) during the 1990 ouburst and 
in 1996, respectively.

\subsection{SAX~J1747.0$-$2853}
The source, observed in outburst by the {\em BeppoSAX}-WFCs in Spring 1998 with a 2--10 keV intensity of 15 mCrab, was 
observed again in outburst in Spring 2000 with a peak 2--10 keV intensity of 140 mCrab by {\em RXTE} and {\em BeppoSAX}.
Taking into account the spectrum observed in 1998 by the {\em BeppoSAX}-NFIs (Natalucci et al., 2000b) and the 
estimated distance of 7.4 kpc, the source luminosity was $\sim 5\times 10^{36} {\rm erg~s}^{-1}$, during the 1998 
outburst and $\sim 5\times 10^{37} {\rm erg~s}^{-1}$ in 2000.
The accretion rate was $4.5\times 10^{-10} M_{\odot} {\rm y}^{-1}$ ($9\times 10^{3} {\rm g~cm}^{-2} {\rm s}^{-1}$).
During the quiescent phase the source was below a $3\sigma$ intensity threshold of 3 mCrab (2--28 keV), leading to upper 
limits of $1.6\times 10^{36} {\rm erg~s}^{-1}$ and $1.5\times 10^{-10} M_{\odot} {\rm y}^{-1}$ ($3\times 10^{3} 
{\rm g~cm}^{-2} {\rm s}^{-1}$) for the luminosity and the accretion rate, respectively.

\subsection{SAX~J1750.8$-$2900}
During the 1997 outburst, the source peaked at 70 mCrab in the 2--28 keV bandpass.  A total of nine type-I X-ray bursts 
were observed by the WFCs, but only the brightest of them showed radius-expansion features (see Fig. 2).  
The evidence for radius expansion is marginal ($\sim 3.5 \sigma$) in the blackbody radius time history, but the high energy 
profile is clearly double-peaked.  
Taking into account the observed spectrum (Natalucci et al., 1999), the peak bolometric luminosity was 
$\sim 5\times 10^{37} {\rm erg~s}^{-1}$, for the inferred distance of 7.5 kpc.  
This corresponds to an accretion value of $4\times 10^{-9} M_{\odot} {\rm y}^{-1}$, i.e. 
$9\times 10^{4} {\rm g~cm}^{-2} {\rm s}^{-1}$.
A $3 \sigma$ upper limit of $\sim 2\times 10^{36} {\rm erg~s}^{-1}$ is obtained for the quiescent emission. In this case, 
the mass transfer rate is $< 2\times 10^{-10} M_{\odot} {\rm y}^{-1}$ ($4\times 10^{3} {\rm g~cm}^{-2} {\rm s}^{-1}$).

\subsection{SAX~J1808.6$-$3658}
This well-known source, the only example of accreting millisecond pulsar so far, was observed in outburst in 1996, 1998 
and 2000.
At the time of its discovery (September 1996, by the {\em BeppoSAX}-WFCs), the source reached a bolometric luminosity 
of $7.5\times 10^{36} {\rm erg~s}^{-1}$ for a distance of 2.8 kpc (in 't Zand et al., 2000).  
In the quiscent phase, the flux was below $\sim 6\times 10^{35} {\rm erg~s}^{-1}$.  These luminosities correspond to 
accretion rates of $7\times 10^{-10} M_{\odot} {\rm y}^{-1}$ ($= 1.4\times 10^{4} {\rm g~cm}^{-2} {\rm s}^{-1}$) and
$< 6\times 10^{-11} M_{\odot} {\rm y}^{-1}$ ($= 1.2\times 10^{3} {\rm g~cm}^{-2} {\rm s}^{-1}$) for the 1996 outburst 
and the quiescent phase, respectively. 

\subsection{SAX~J1810.8$-$2609}
This weak transient, discovered by the {\em BeppoSAX}-WFCs during an outburst in March 1998, reached a 2--28 keV 
intensity of 15 mCrab.  On the basis of the observed spectrum (Natalucci et al., 2000a) and for the distance of 
$\sim 5$ kpc obtained via a single WFC-detected Eddington-limited burst, a bolometric luminosity of 
$3.5\times 10^{36} {\rm erg~s}^{-1}$ is derived, corresponding to an average accretion rate of 
$3\times 10^{-10} M_{\odot} {\rm y}^{-1}$ ($= 6\times 10^{3} {\rm g~cm}^{-2} {\rm s}^{-1}$).
Upper limits of $\sim 7\times 10^{35} {\rm erg~s}^{-1}$ and $6\times 10^{-11} M_{\odot} {\rm y}^{-1}$ 
($= 1.2\times 10^{3} {\rm g~cm}^{-2} {\rm s}^{-1}$) are obtained in the quiescent phase for the luminosity and the 
mass transfer rate, respectively.

\subsection{4U~1812$-$12}
4U~1812$-$12 is a weak pesistent source with average 2--10 keV intensity of $\sim 4\times 10^{-10} {\rm erg~cm}^{-2} {\rm s}^{-1}$
(e.g. Warwick et al., 1981).
For a distance value of 3.4 kpc, we extrapolate a luminosity of $4\times 10^{36} {\rm erg~s}^{-1}$ (see also 
Cocchi et al., 2000a).  Again, for standard neutron star parameters the accretion rate is then 
$3.5\times 10^{-10} M_{\odot} {\rm y}^{-1}$, corresponding to $7\times 10^{3} {\rm g~cm}^{-2} {\rm s}^{-1}$.

\section*{ACKNOWLEDGMENTS}
We thank the staff of the {\em BeppoSAX Science Operation Centre} and {\em Science
Data Centre} for their help in carrying out and processing the WFC Galactic Centre
observations. The {\em BeppoSAX} satellite is a joint Italian and Dutch program.
M.C., A.B., L.N. and P.U. thank Agenzia Spaziale Nazionale ({\em ASI}) for grant support.


\begin{thebibliography}{}


\bibitem[Cocchi et al. (1999a)]{cocc99a}
Cocchi M., et al., 1999a, IAUC 7247

\bibitem[Cocchi et al. (1999b)]{cocc99b}
Cocchi M., et al., 1999b, proc. 5th Compton Symp. 1999, in press

\bibitem[Cocchi et al. (1999c)]{cocc99c}
Cocchi M., et al., 1999c, A\&A 346, L45

\bibitem[Cocchi et al. (2000a)]{cocc00a}
Cocchi M., et al., 2000a, A\&A 357, 527

\bibitem[Cocchi et al. (2000b)]{cocc00b}
Cocchi M., et al., 2000b, ApJ, submitted

\bibitem[Heise et al. (1999)]{heis99}
Heise J., et al., 1999, Ap. Lett. \& Comm. 38/1-6, 297

\bibitem[Hoffman et al. (1978)]{hoff78}
Hoffman J.A., Marshall H.L. \& Lewin W.H.G., 1978, Nature 271, 630

\bibitem[in 't Zand et al. (1998)]{zand98}
in 't Zand J.J.M., et al. 1998, A\&A 331, L25

\bibitem[in 't Zand et al. (1999)]{zand99}
in 't Zand J.J.M., et al. 1999, IAUC 7243

\bibitem[in 't Zand et al. (2000)]{zand00}
in t Zand J.J.M., et al. 2000, A\&A submitted

\bibitem[Jager et al. (1997)]{jage97}
Jager R., et al., 1997, A\&A, 125, 557

\bibitem[Kaptein et al. (2000)]{kapt00}
Kaptein R.G., et al., 2000, A\&A, 358, L71

\bibitem[Lewin et al. (1984)]{lewi84}
Lewin W.H.G., Vacca W.D. \& Basinska E.M., 1984, ApJ 277, L57

\bibitem[Lewin et al. (1993)]{lewi93}
Lewin W.H.G., van Paradijs J. \& Taam R.E., 1993, Sp. Sci. Rev. 62, 223

\bibitem[Nakamura et al. (1989)]{naka89}
Nakamura N., et al., 1989, PASJ 41, 617

\bibitem[Natalucci et al. (1999)]{nata99}
Natalucci L., et al., 1999, ApJ 523, L45

\bibitem[Natalucci et al. (2000a)]{nata00a}
Natalucci L., et al., 2000a, ApJ 536, 891

\bibitem[Natalucci et al. (2000b)]{nata00b}
Natalucci L., et al., 2000b, ApJL in press

\bibitem[Olive et al. (2000)]{oliv00}
Olive J.F., et al., 2000, Adv. Sp. Res. 25/3-4, 369

\bibitem[Pavlinsky et al. (1994)]{pavl94}
Pavlinsky M.N., Grebenev S.A. \& Sunyaev R.A., 1994, ApJ 425, 110

\bibitem[Smith et al. (1999)]{smit99}
Smith M.J.S., et al., 1999, Ap. Lett. \& Comm. 38/1-6, 137

\bibitem[Tawara et al. (1984)]{tawa84}
Tawara Y., et al., 1984, ApJ 276, L41

\bibitem[Ubertini et al. (1999)]{uber99}
Ubertini P., et al., 1999, Ap. Lett. \& Comm. 38/1-6, 301

\bibitem[van Paradijs \& Mc Clintok (1994)]{vanp94}
van Paradijs J., and Mc Clintok J.E., 1994, A\&A, 290, 133


\bibitem[Warwick et al. (1981)]{warw81}
Warwick R.S., et al., 1981, MNRAS 197, 865

\end{thebibliography}
\end{document}